# The Updated Genome Warehouse: Enhancing Data Value, Security, and Usability to Address Data Expansion

Yingke Ma[1,2,#], Xuetong Zhao[1,2,#], Yaokai Jia[1,2,#,†], Zhenxian Han[1,2,#], Caixia Yu[1,2], Zhuojing Fan[1,2], Zhang Zhang[1,2,3], Jingfa Xiao[1,2,3], Wenming Zhao[1,2,3], Yiming Bao[1,2,3,*], Meili Chen[1,2,3,*]

[1] National Genomics Data Center, China National Center for Bioinformation, Beijing 100101, China

[2] Beijing Institute of Genomics, Chinese Academy of Sciences, Beijing 100101, China

[3] University of Chinese Academy of Sciences, Beijing 100049, China

[#] Equal contribution.

[*] Corresponding authors.

E-mail: chenml@big.ac.cn (Chen M), baoym@big.ac.cn (Bao Y).

[†] Current address: AVIC China Aero-Polytechnology Establishment, Beijing 100028, China

**Running title**: *Ma Y et al / Genome Assembly Repository in Updated GWH*

Words (2596), references (13), figures (2), tables (1), supplementary figures (0) and supplementary tables (0).

Title (88), running title (44), keyword (5), abstract (134).

**Abstract**

The Genome Warehouse (GWH), accessible at https://ngdc.cncb.ac.cn/gwh, is an extensively utilized public repository dedicated to the deposition, management and sharing of genome assembly sequences, annotations, and metadata. This paper highlights noteworthy enhancements to the GWH since the 2021 version, emphasizing substantial advancements in web interfaces for data submission, database functionality updates, and resource integration. Key updates include the reannotation of released prokaryotic genomes, mirroring of genome resources from National Center for Biotechnology Information (NCBI) GenBank and RefSeq, integration of *Poxviridae* sequences, implementation of an online batch submission system, enhancements to the quality control system, advanced search capabilities, and the introduction of a controlled-access mechanism for human genome data. These improvements collectively augment the ease and security of data submission and access as well as genome data value, thereby fostering heightened convenience and utility for researchers in the genomic field.

# Introduction

Genome Warehouse (GWH, https://ngdc.cncb.ac.cn/gwh) [1], one of the major resources in National Genomics Data Center (NGDC, https://ngdc.cncb.ac.cn) [2] of China National Center for Bioinformation (CNCB, https://www.cncb.ac.cn) [3], serves as a publicly accessible repository of genome assembly data. Since its inception in 2017, GWH has been accepting genome assembly sequences, annotations, and metadata submissions from researchers worldwide. Comparing to the well-known international genome databases such as National Center for Biotechnology Information (NCBI) WGS [4] and European Bioinformatics Institute (EBI) Ensembl [5], which were established in early 2000's with high data volume and significant influence, GWH's data resources and service system are in relatively smaller scale and less efficient, but continuously improving and gradually catching up. It has been widely recognized by the scientific community, as testified by increasing users, downloads, and supported journals. As of Nov. 18, 2024, GWH has accepted 84,660 genome assemblies that are submitted by 1086 submitters from 363 institutions, reported by 591 research articles in 161 scientific journals for released genome assemblies. However, the limited storage, network bandwidth, staff members and funding remain limiting factors for GWH's further development.

In the era of technology evolution and heightened user expectations for improved data services, GWH has encountered additional challenges. First, the tension between the burgeoning growth of genomic data and the efficiency of data submission is becoming increasingly prominent [6]. Second, despite the stable operation of the in-house quality control system, the growing data complexities present significant hurdles to the meticulous quality control and management of genomic data submissions. Third, since the genome annotations serve as the bridge between genome sequence and inferred biological function, the omission of annotation significantly diminishes the potential value of genome data utilization and hinders the realization of the full spectrum of data-driven insights and discoveries. And finally, the inability to search and access the International Nucleotide Sequence Database Collaboration (INSDC) [7] data through GWH contributes to data fragmentation and the creation of data silos.

In addressing the challenges presented by the exponential expansion of data, the imperative of ensuring data quality, optimizing data utilization value, and facilitating streamlined data access for archival and management purposes, we introduce here an updated implementation of GWH. These enhancements mainly include the reannotation of prokaryotic genomes, mirroring of data from NCBI [4], introduction of online batch submission pipeline, enhancements to data quality control, refined management of controlled-access to human genome data, and integration of advanced search functionalities. These improvements significantly elevated user experience and position GWH as a more efficient complement to the INSDC.

## Updates overview and data statistics

This paper provides a comprehensive overview of these major developments. In terms of data resources, significant updates encompass prokaryotic genome reannotation and NCBI data integration. Concerning functionality, key updates include batch submission, quality control enhancements, advanced search capabilities and the introduction of controlled-access systems. Table 1 provides a detailed enumeration of all the updates.

As of Nov. 18, 2024, the submission volume of GWH has showed remarkable growth, with a total of 84,660 genome assembly submissions accepted. These submissions contain 4804 organisms from a diverse community of 1086 submitters, representing 363 organizations. Notably, 24 international submitters from 5 countries have contributed to this global initiative. GWH has successfully released 57,686 genome assemblies, representing 3697 distinct organisms across various assembly levels (Figure 1). The released genome assemblies span diverse divisions, including animals (3017), plants (2971), fungi (235), protists (951), bacteria (4348), archaea (127), viruses (2456), metagenomes (43,165), and others (416). These valuable genomic resources have been disseminated through 591 articles published in 161 journals.

Moreover, GWH has received 62 access requests for controlled human genome data, with 16 of them being authorized by the data provider. The GWH website has garnered attention from a global audience, attracted 214,182 unique users spanning 170 countries/regions. From 2021 to 2024, daily download for released genome assemblies consistently surpassed 500 times. Recognizing its dedication to ongoing maintenance and updates, GWH has received certifications from FAIRsharing.org and re3data.org, validating its adherence to standards for findability, accessibility, interoperability, and reusability. In summary, this diverse statistical metrics collectively illustrate the substantial growth of GWH, emphasizing its significant potential to support global research efforts in genomics.

## Data resource updates

### Prokaryotic genome reannotation

Among the 13,919 prokaryotic genome submissions to GWH, less than 0.5% are accompanied by genome annotations. Moreover, the submitted genome annotations primarily contain gene structure predictions, lacking comprehensive gene function annotations. To enhance the utility of released prokaryotic genomes for associated research endeavors, this version of GWH introduces an automatic reannotation pipeline utilizing the Prokaryotic Genome Annotation Pipeline (PGAP) [8] from NCBI [4]. This initiative aims to provide standardized and unified genome reannotations, thereby facilitating progress in functional genomics research. The reannotation pipeline involves the following six processes (Figure 2): (1) Genome size validation: Genomes falling outside the expected genome size range or that have genome size over 30 mega-base pairs (Mbp) or less than 1 Mbp are skipped. (2) Input preparation: Topology and location information (affects genetic codon selection) in genome sequence files are integrated into the definition line if known. (3) Taxon name check: The taxon name is crucial as it influences genetic codes and reference proteins selection. To ensure completeness and precision, the automatic pipeline incorporates a PGAP taxon check, which confirms or corrects the taxon name based on the average nucleotide identity (ANI) between the query genome and the type genomes stored in GenBank [9]. If misassigned check result occurs, the taxon name is corrected to align with the best-matched taxon name before continuing with PGAP annotation. If contamination is identified, the reannotation procedure is terminated. Additionally, PGAP requires the submitted taxon rank to be at least at the genus level; hence, a hypothetical taxon name is employed for taxon correction if the submitted taxon is above genus. Only genomes with confirmed results or those with misassigned results exhibiting a high confidence level undergo PGAP annotation. (4) PGAP annotation: The annotation result contains structural rRNA, tRNA, small ncRNA, repetitive regions, protein-coding genes, and protein naming. (5) Archival: Upon completion of PGAP, the pipeline assigns unique accession numbers, generates downloadable files, and implements comprehensive backup measures. (6) Release: Subsequently, the reannotation results are released and displayed on the page of original submitted genome sequences as associated data pairs, readily browsable, retrievable, and downloadable. These files are publicly accessible at https://download.cncb.ac.cn/assembly/gwh/reannotation. As of Nov. 18, 2024, GWH has completed annotation for 3986 out of 4475 released prokaryotic genomes. The remaining genomes exhibit abnormal sizes (170), lack taxonomy rank at genus/species levels and cannot be corrected (312), and contain contamination (7).

**NCBI data integration**

To enhance data sharing and ensure comprehensive accessibility to all genome assembly-related information, GWH has implemented an in-house mirroring pipeline. This pipeline mirrors the entire genome assembly data from GenBank [9] and RefSeq [10], capturing metadata primarily through NCBI E-utilities, genome sequences. annotation files, and corresponding statistical data sourced from NCBI FTP files (https://ftp.ncbi.nlm.nih.gov/genomes/genbank/ and https://ftp.ncbi.nlm.nih.gov/genomes/refseq/). Additionally, GWH retrieves taxon information from the NCBI taxonomy database, including scientific names, common names, synonymous names, and taxonomy lineages. This extensive dataset seamlessly integrates with data directly submitted to GWH. The platform ensures continuous synchronization, updates, and comprehensive support for searching, browsing, and downloading data (https://download.cncb.ac.cn/assembly/ncbi/). As of Nov. 18, 2024,

GWH has successfully mirrored approximately 2.51 million GenBank and 0.47 million RefSeq genome assembly records, along with associated data files. The all-in-one integration of genome data enhances researchers' convenience in accessing NCBI data (https://ngdc.cncb.ac.cn/gwh/browse/assembly?source=ncbi).

In our effort to integrate virus genome assembly data, we aggregate genomic and protein sequences, along with associated metadata related to the *Poxviridae* family sourced from NCBI, to establish the poxvirus module (https://ngdc.cncb.ac.cn/gwh/browse/virus/poxviridae) [11]. The resource undergoes daily updates by an in-house script, ensuring timely information availability. Users are able to browse and filter sequences based on diverse criteria, facilitating efficient data exploration and retrieval.

## Function updates

### Data submission and quality control

To enhance user experience with a focus on friendliness, convenience, and efficiency in data submission, the latest version of GWH introduces the online batch submission functionality. This feature empowers submitters to submit all genome assembly data associated with the same BioProject and publication simultaneously. Users can provide unique metadata for each assembly, such as BioSample accession, assembly name, assembly level, and genome sequence filename, using an Excel template file. Subsequent to the submission of metadata and data files, the entire batch undergoes a comprehensive quality control and feedback process. This streamlined approach eliminates the need for repetitive submission of common metadata, significantly reducing the submission process duration and markedly enhancing submission efficiency.

High-quality genomic data forms a crucial foundation for accurate and reliable downstream omics data analysis. Recognizing this, GWH has undertaken a comprehensive update to its quality control system to ensure the submitted data quality. These enhancements consist of four key components: (1) Genome size check: GWH cross-references genome size data with information from NCBI

(ftp://ftp.ncbi.nlm.nih.gov/genomes/ASSEMBLY_REPORTS/species_genome_size.txt.gz) to verify a reasonable genome size. (2) Lineage rank check: GWH requires a genus/species level description for each assembly, ensuring taxonomic clarity. (3) Enhanced precision in genome annotation validation: Building upon the original quality control system, GWH integrates NCBI's table2asn tool (https://www.ncbi.nlm.nih.gov/genbank/table2asn/) to validate genome annotations using stricter criteria for gene structure and function annotations, as well as the file format, completeness, and consistency of gene structures. (4) Warnings. In addition to identifying fatal errors, GWH now issues warnings, allowing submitters to consider modifications based on biological context of their data. With over 800 different types of error/warning messages, GWH expands its quality control capabilities to ensure rigorous scrutiny of genome annotation content.

Furthermore, GWH has refined the accepted data types, placing a particular emphasis on comprehensive eukaryotic and prokaryotic genomes, as well as metagenomes. Given the rapid expansion of metagenomics data, which represents the collective genomes of microorganisms within a specific environment, the availability of rich metadata and accurate genomic data is essential for facilitating scientific exploration of microbial community composition and function. To support this endeavor, GWH has enhanced its data organization structure and substantially augmented the collection of metadata specifically related to metagenome-assembled genomes (MAGs). This update ensures seamless linkage among sequencing raw data, primary metagenome assembly, binned metagenome, and MAGs. Furthermore, this version incorporates essential metadata pertaining to binning analysis, quality assessment, and taxonomic identification for both binned metagenomes and MAGs. For example, completeness, contamination rate, binning analysis software, and associated parameters. By including rich metadata and a well-organized data structure, GWH allows users to select and utilize metagenome-associated data more efficiently, thereby enhancing the scientific exploration of microbial communities.

As sequencing technology advances, an increasing number of new organisms undergo sequencing and assembly. However, the taxonomy information for these species may not be readily available in the NCBI Taxonomy database, hindering the sharing process for their genomes. To address this challenge, GWH allows users to contribute the pertinent taxonomy information of newly sequenced organisms with supporting evidence from publications or other databases such as GTDB [12]. Currently, GWH has incorporated data for 1382 new organisms, comprising 1537 genome assemblies. Furthermore, the genetic code determines the codon usage for each coding sequence (CDS) within an organism. Notably, some GWH submitters have identified inaccuracies in the genetic code for lesser-known organisms in NCBI's Taxonomy database and have provided corrected genetic code values following thorough analysis. Warning messages about these exceptions are prominently displayed on the detailed web page for each assembly. GWH currently features nine organisms across 18 genome assemblies, employing these updated genetic codes. These data primarily stem from the genomes of protists contributed by the P10K project [13], offering insights into substantial genetic code variations among ciliates.

**Data retrieval, search and access**

The latest version of GWH introduces significant enhancements to data retrieval, download, and access. GWH now provides personalized data retrieval services through the introduction of an advanced filtering function on the assembly browse page. Users can tailor their searches by setting single or multiple filter conditions, such as scientific name, reannotation status, and GC content range, enabling convenient identification and retrieval assemblies of interest. To enhance the convenience of data download for GWH users, we have introduced a batch download function for filtered results. This feature is applicable to both GWH assemblies/annotations and integrated NCBI assemblies/annotations. Additionally, GWH has incorporated the HTTPS download method to adapt to the evolving landscape of web browsers. These updates are designed to improve the efficiency and user-friendliness of data retrieval and download services within GWH.

Furthermore, GWH has introduced an advanced search system with 18 conditions (*e.g.*, BioProject, BioSample, assembly level, and organism group) for all released genome assemblies, including those integrated from GenBank and RefSeq. This robust system supports both single and multi-condition queries, allowing users to specify conditions for particular fields or across all fields. Importantly, users can retrieve genomes of corresponding organisms from any rank in the lineage, adding flexibility to their search conditions. The browsing session retains a search history, enabling users to incorporate each historical search as a condition in the search expression. On the result page, entries are categorized and aggregated, with additional multi-dimensional filtering options provided to further customize search results. Users can select metadata and data download links of interest from the search results, and download them as a text file. Alternatively, batch download function is provided for the metadata and data files download for resulting genome assemblies to the convenience of GWH users. The advanced search function empowers users to construct detailed search formulas, facilitating precise data searches and ensuring that the results are concise and accurate. This enhancement greatly enhances the usability and efficiency of the search functionality within GWH.

To meet the unique management needs of human genetic resources, we have implemented a controlled-access data request and authorization management system. Access to controlled data is restricted to users who have obtained explicit permission from the data owner, and the downloaded data can only be used within its designated validity period. Furthermore, users are obligated to adhere to legal and compliance standards, conducting research solely within the declared scope. This system supports online interactive operations and ensures prompt email notifications, enabling efficient communications between applicants and data owners throughout the application and processing stages. It fosters efficient collaboration and contributes to the protection of sensitive information and responsible utilization of controlled data.

## Future directions

GWH serves as an accessible and user-friendly platform dedicated to archiving, searching, and sharing genome data, complemented by a range of associated services. However, the accelerating influx of genome data into GWH presents significant challenges in processing and managing this vast dataset. Therefore, GWH remains committed to continuous refinement of its data curation platform, with a focus on enhancing automation and intelligence across all processes. As genomic research progresses, GWH aims to broaden its accepted data types and reinforce quality control measures, including the incorporation of pan-genome data in gfa format. Additionally, GWH plans to explore the integration of eukaryotic genome annotation services to enhance the value of submitted genome data and promote advancements in related research fields. This initiative includes providing comprehensive gene function annotations, GO functions, metabolic pathways, and protein domains. GWH will utilize the open-source NCBI Eukaryotic Genome Annotation Pipeline - External (EGAPx, https://github.com/ncbi/egapx), the pipeline used by RefSeq, to establish its eukaryotic reannotation pipeline for supported organisms. For other organism, GWH will employ an optimized pipeline developed in collaboration with our partner institute to ensure comprehensive genome reannotations. Looking ahead, GWH envisions facilitating the sharing and exchange of all publicly available genome assembly data with INSDC members. By doing so, GWH seeks to contribute to a global collaborative effort and provide researchers worldwide with comprehensive and enriched genomic data resources, thereby advancing scientific discovery in genomics.

## Data availability

GWH is freely accessible at https://ngdc.cncb.ac.cn/gwh/.

## Authors' contributions

Yingke Ma: Software, Writing - original draft, review & editing. Xuetong Zhao: Investigation, Methodology, Data curation, Writing - original draft. Yaokai Jia: Software, Writing - original draft. Zhenxian Han: Software. Caixia Yu: Data curation. Zhuojing Fan: Methodology. Zhang Zhang: Writing - review & editing. Jingfa Xiao: Writing - review & editing. Wenming Zhao: Writing - review & editing. Yiming Bao:

Conceptualization, Writing - review & editing, Supervision. Meili Chen: Project administration, Investigation, Methodology, Data curation, Writing - original draft, review & editing. All authors have read and approved the final manuscript.

## Competing interests

The authors have declared no competing interests.

## Acknowledgements

We thank the hardware and system administration members from NGDC, especially Yubin Sun and Shuang Zhai, for their help in server configuration and maintenance. This work was supported by the Science & Technology Fundamental Resources Investigation Program [Grant No. 2022FY101203 to XZ], the International Partnership Program of the Chinese Academy of Sciences [Grant No. 161GJHZ2022002MI to YB], the Professional Association of the Alliance of International Science Organizations [Grant No. ANSO-PA-2023-07 to YB], the CAS Key Technology Talent Program to MC, the Open Biodiversity and Health Big Data Programme of IUBS to YB, National Key Research & Development Program of China [Grant No. 2023YFC2605700 to WZ], and the National Natural Science Foundation of China [Grant Nos. 32170678 to WZ, 32170669 to JX, 32030021 to ZZ].

## ORCID

0000-0002-9460-4117 (Yingke Ma)
0000-0002-3019-8615 (Xuetong Zhao)
0009-0003-9677-4624 (Yaokai Jia)
0009-0001-4467-5080 (Zhenxian Han)
0000-0002-3882-9979 (Caixia Yu)
0009-0003-3575-6967 (Zhuojing Fan)

0000-0001-6603-5060 (Zhang Zhang)

0000-0002-2835-4340 (Jingfa Xiao)

0000-0002-4396-8287 (Wenming Zhao)

0000-0002-9922-9723 (Yiming Bao)

0000-0003-0102-0292 (Meili Chen)

# Table and figure legends

**Table 1. Comparison of GWH between the two versions in 2024 and 2021**

| Category | 2024 | 2021 |
|---|---|---|
| **Prokaryotic genome reannotation** | Available | / |
| **NCBI genome assembly resource** | NCBI GenBank and RefSeq genome assemblies | / |
| **Poxvirus sequences resource** | Available | / |
| **Data submission** | Online batch submission | Online single submission and offline batch submission |
| **Accepted genome data** | Full genomes of eukaryote and prokaryote and metagenomes | Assembled virus, eukaryote, prokaryote, organelle, and plasmid genomes and metagenomes |
| **Quality control** | NCBI table2asn and updated in-house script | In-house script |
| **Search function** | BIGSearch and advanced search | BIGSearch |
| **Controlled access** | Controlled human genome data can be downloaded upon request | / |
| **Data download** | FTP and HTTPS; online batch download | FTP |
| **Language** | English and Chinese | English |
| **BLAST** | Available | / |
| **Data statistics** | 11 modules | 8 modules |

**Figure 1. Direct submitted genome assembly data statistics in GWH**

(A) All direct submitted assembly data; (B) Publicly released assemblies. These statistical data are

as of Nov. 18, 2024.

**Figure 2. The workflow of prokaryotic genome reannotation in GWH**

The process of prokaryotic genome reannotation in GWH involves preprocessing and annotation of genome sequences, annotation result archival and release of annotation results.

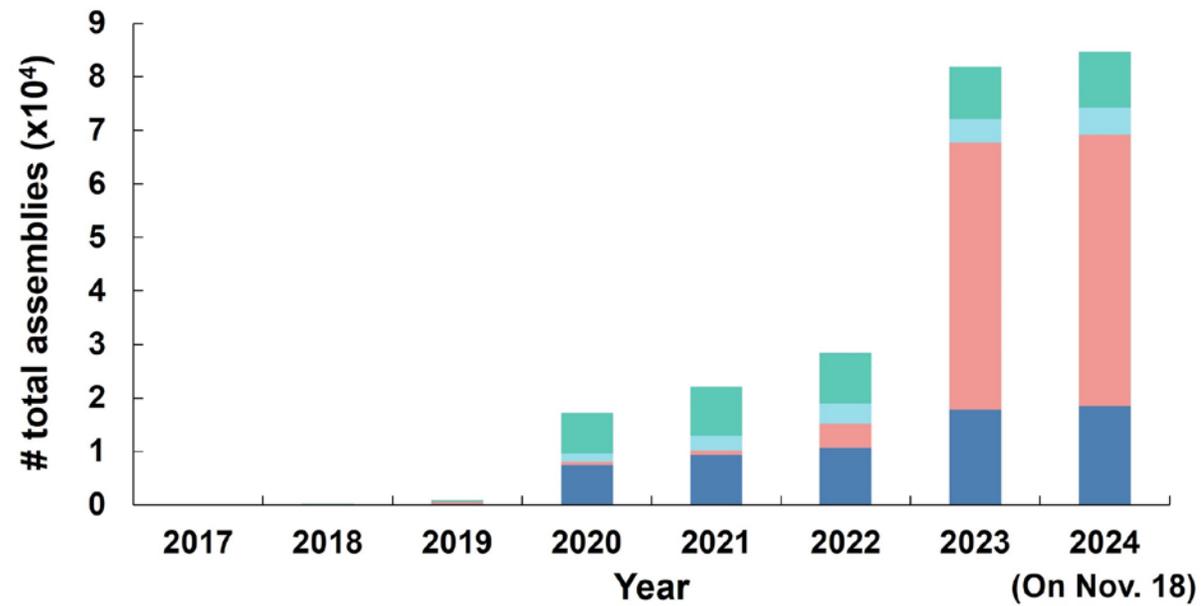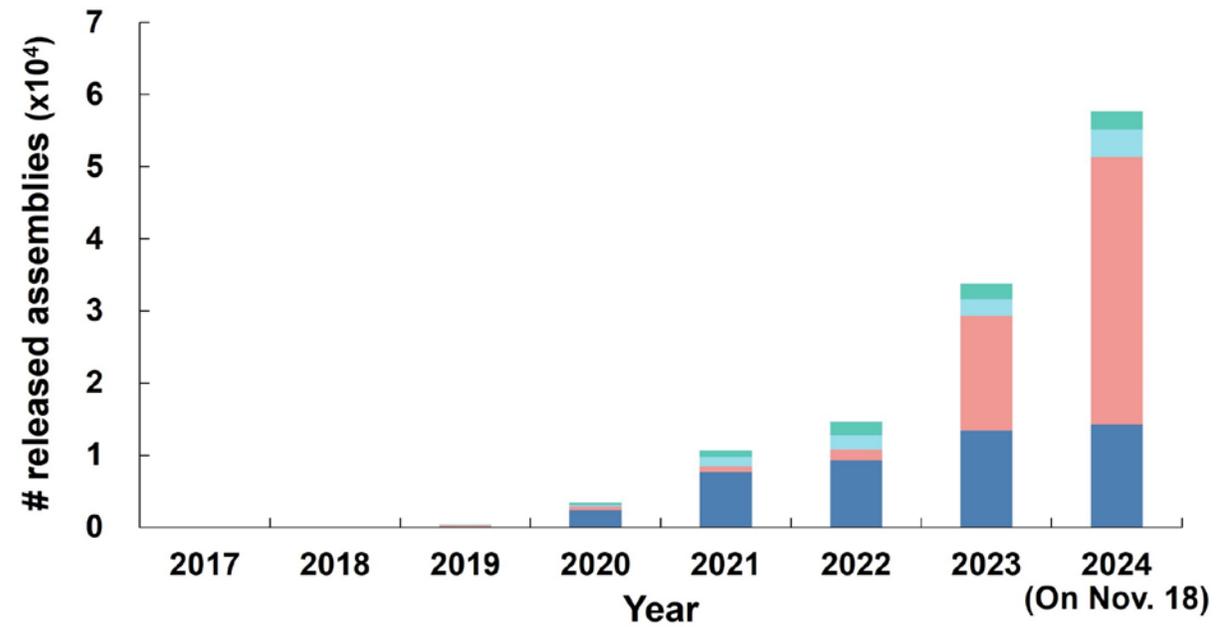

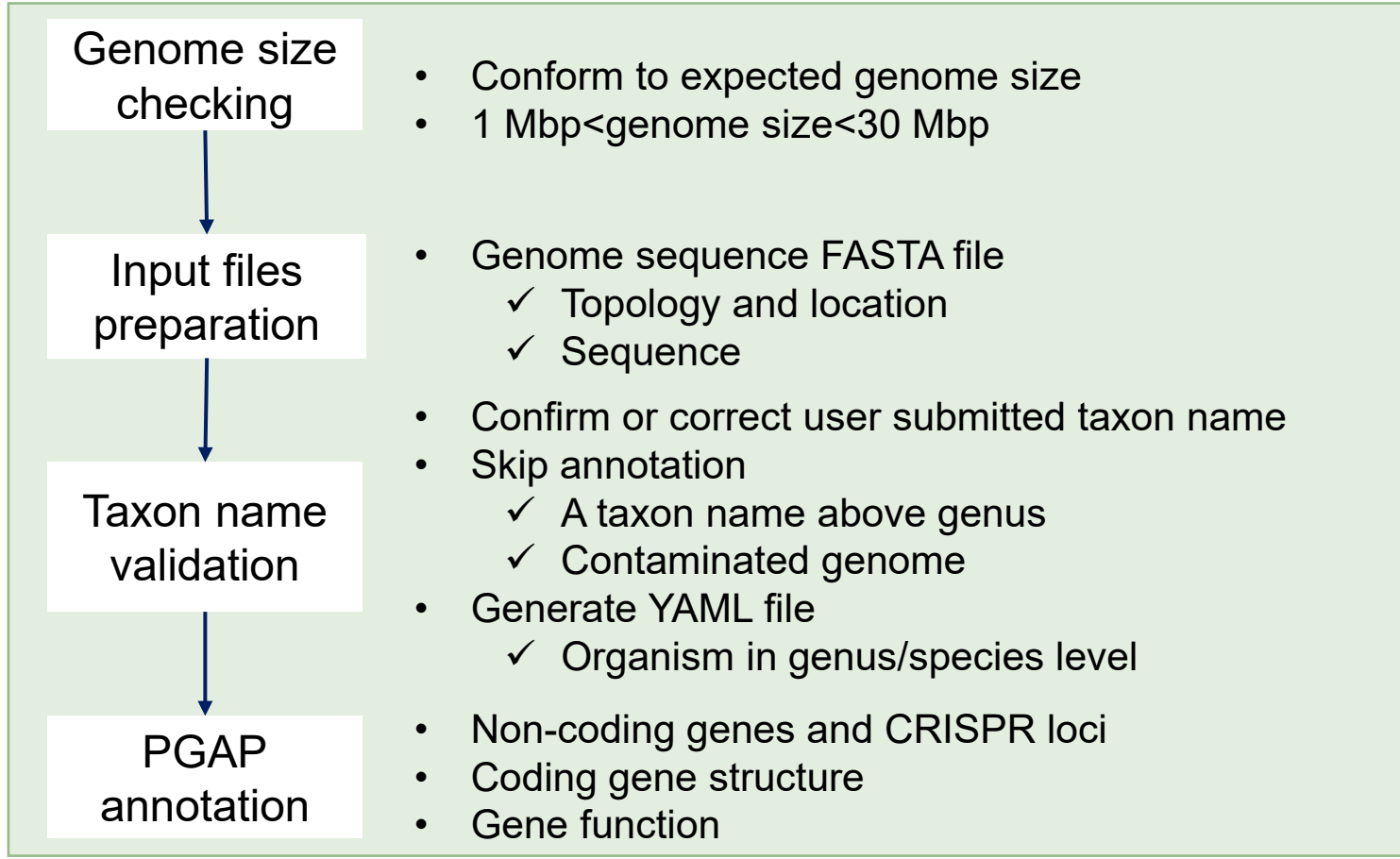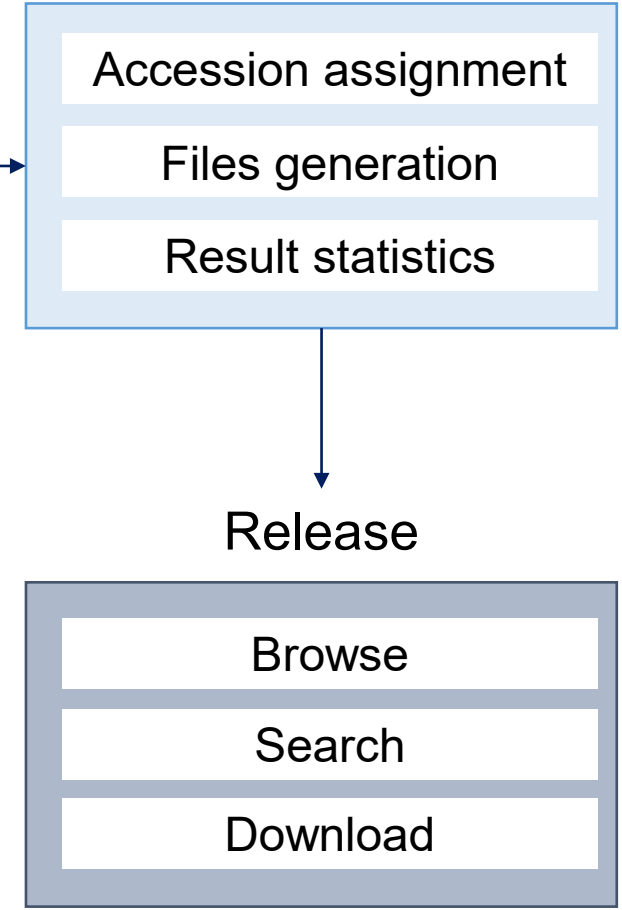